\documentclass[draft]{agujournal2019}
\usepackage{url} 
\usepackage{lineno}
\usepackage{soul}
\usepackage{amsmath}
\usepackage{ragged2e}             

\draftfalse
\journalname{Geophysical Research Letters}

\begin{document}

\title{Why are Mountaintops Cold? The Transition of Surface Lapse Rate on Dry Planets}

\authors{Bowen Fan\affil{1}, Malte F. Jansen\affil{1}, Michael A. Mischna\affil{2}, Edwin S. Kite\affil{1}}

\affiliation{1}{Department of the Geophysical Sciences, University of Chicago, Chicago, IL, 60637, USA}
\affiliation{2}{Jet Propulsion Laboratory, California Institute of Technology, Pasadena, CA, 91109, USA}

\correspondingauthor{Bowen Fan}{bowen27@uchicago.edu}

\begin{keypoints}
\justifying
\item Surface lapse rate robustly increases with atmospheric longwave optical thickness (greenhouse effect) in a general circulation model.
\justifying
\item Increased pressure further contributes to a tropical surface lapse rate increase in moderately opaque atmospheres.
\justifying
\item A simple model, assuming weak temperature gradient and highland convective adjustment, provides insight into the mechanisms.

\end{keypoints}

\begin{abstract}
\justifying
Understanding surface temperature is important for habitability. Recent work on Mars has found that the dependence of surface temperature on elevation (surface lapse rate) converges to zero in the limit of a thin $\mathrm{CO_2}$ atmosphere. However, the mechanisms that control the surface lapse rate are still not fully understood. It remains unclear how the surface lapse rate depends on both greenhouse effect and surface pressure. Here, we use climate models to study when and why ``mountaintops are cold". We find the tropical surface lapse rate increases with the greenhouse effect and with surface pressure. The greenhouse effect dominates the surface lapse rate transition and is robust across latitudes. The pressure effect is important at low latitudes in moderately opaque ($\tau \sim 0.1$) atmospheres. A simple model provides insights into the mechanisms of the transition. Our results suggest that topographic cold-trapping may be important for the climate of arid planets.
\end{abstract}

\section*{Plain Language Summary}
\justifying
Understanding surface temperature on a planet is important for life on Earth and beyond. On Earth, we know ``mountaintops are cold", which means that surface temperature decreases with elevation. However, this idea does not apply on present-day Mars. Here, we investigate when and why the Earth-based understanding holds for planets with different types of atmospheres. Using a global climate model, we show that both the greenhouse effect (atmospheric infrared opacity) and the pressure effect (atmospheric turbulence) are important. The weaker the greenhouse effect, or the thinner the atmosphere, the slower the surface cools with elevation. The greenhouse effect plays the dominant role, but in moderately opaque atmospheres, the pressure effect becomes important as well. Our work reveals a novel connection between climate and geomorphology. For example, on a planet with a relative thin pure $\mathrm{O_2}$ or $\mathrm{N_2}$ atmosphere, we do not expect that ``mountaintops are cold".

\section{Introduction}
\justifying
Surface temperature, $T_s$, is fundamental for understanding habitability \cite{Seager2013}. In addition to the direct implications of surface temperature for life, the distribution of surface temperature defines the cold trap (where moisture tends to condense and accumulate), which regulates the hydrological cycle on an arid planet \cite{mitchell2016,Ding2020}. 

\justifying
There are two paradigms for estimating the distribution of surface temperature. The first paradigm, which we refer to as ``radiation deficits are cold", states that the coldest region is close to the time-mean minimum of solar radiation. These regions are the deficits of net radiation flux at the top of the atmosphere. For example, the poles are regions of radiation deficit for Earth and modern Mars, the night hemisphere is the region of radiation deficit for synchronously rotating exoplanets \cite{wordsworth2015atmospheric}, and the tropics were the region of radiation deficit for pre-modern Mars at times when the obliquity was very high \cite{forget2006}. This paradigm focuses on the large-scale pattern of surface temperature, and has been extensively studied \cite<e.g.,>{held1993,forget2006,kaspi2015,wordsworth2015atmospheric}. 

\justifying
The second paradigm, which we refer to as ``mountaintops are cold," emphasizes local processes and emerged long before the modern era. This paradigm states that the change of surface temperature with elevation should follow that of the atmosphere, which can be quantified as the lapse rate. For a well-mixed, isolated atmospheric column without condensible species, the lapse rate is the dry adiabat, $\Gamma_{ad}$:

\begin{equation}
    \Gamma_{ad} = -\frac{dT_a}{dZ} = \frac{g}{c_p} 
\end{equation}
where $T_a$ is atmospheric temperature, $Z$ is height, $g$ is gravity, and $c_p$ is the specific heat of air. If surface temperature also follows this adiabatic lapse rate, we therefore expect colder temperature at higher elevations. Taken together, these two paradigms predict how surface temperature changes horizontally (by solar radiation) and following surface elevation (by gravity and atmospheric composition).

\justifying
However, the idea that ``mountaintops are cold" does not apply to Mars today. Surface lapse rate, $\Gamma_s$, (the change of surface temperature with elevation) is weak on modern Mars \cite{Sagan1968}. Recent work on early Mars has linked the Martian surface lapse rate to the atmosphere's evolution \cite{forget2013,wordsworth2013,Wordsworth2016,Kite2019}. Specifically, $\Gamma_s$ is close to $\Gamma_{ad}$ only for scenarios with thick $\mathrm{CO_2}$ atmospheres. For thin $\mathrm{CO_2}$ atmospheres, $\Gamma_{s}$ is close to zero. From the perspective of surface energy budgets, \citeA{forget2013} and \citeA{Wordsworth2016} suggested that both the sensible heat flux, $SH$, and atmospheric longwave heating, $LW_a$, are important in modulating $T_s$ when the $\mathrm{CO_2}$ atmosphere goes from thick to thin. However, the mechanisms that control the transition in $\Gamma_s$ are not fully understood. It remains unclear whether $SH$ or $LW_a$ is more important for the change of $\Gamma_s$ seen in earlier work, and what controls the changes in $SH$ and $LW_a$. Relatedly, it is not clear in how far the surface lapse rate is sensitive primarily to the change in the surface pressure, versus the change in the greenhouse effect. While the surface pressure and greenhouse effect are directly linked in a pure $\mathrm{CO_2}$ atmosphere, understanding the role of these two distinct effects is important to predict the surface lapse rate on planets with different atmospheres. Are mountaintops still cold for planets with, for example, a thick, pure $\mathrm{O_2}$ atmosphere (high pressure, weak greenhouse effect)? What about a thin, fluoride atmosphere (e.g., $\mathrm{CF_4}$ or $\mathrm{SF_6}$, low pressure, strong greenhouse effect), as suggested by \citeA{marinova2005}?

\justifying
In this paper, we seek a better understanding of when and why ``mountaintops are cold", for planets with different greenhouse gas forcings and atmospheric pressures. Following \citeA{koll2016}, we focus on ``dry planets" (idealized planets forced by gray radiation) to gain a basic understanding of the phenomenon. We introduce our methodology in Section~2. We present and analyse the results in Section~3. Section~4 includes our conclusion, limitations of this research, and implications for future work on different planets.

\section{Methods}
\subsection{General Circulation Model (GCM)}
\justifying
We use the \mbox{MarsWRF} GCM \cite{Richardson2007,Toigo2012} to investigate temperature distribution across different atmospheres. The model resolution is 72$\times$36$\times$40 gridpoints in longitude/latitude/height. All simulations are run for 20 years with 5 years of spin up and averages taken over the last 15 years. 

\justifying
To aid understanding, we use idealized simulations with the following settings. The radiative transfer is computed using a gray gas scheme. Under the scheme, the longwave absorption coefficient, $\kappa$, is varied, allowing us to decouple the greenhouse effect from surface pressure. The shortwave Rayleigh scattering and absorption are set to zero. Surface albedo is uniformly zero. We also carry out simulations with a pure $\mathrm{CO_2}$ atmosphere, using a correlated-k scheme for radiative transfer \cite{Mischna2012} to validate our simulations against earlier studies (Supplementary Information~A). 

\justifying
For the default simulations, the planetary obliquity and orbital eccentricity are set to zero, with solar constant 75$\%$ of the modern Martian value, representing the faint young Sun. Diurnal cycles are disabled. Planetary size and rotation rate are set to Mars values. We explore different planetary climates over a 2D parameter space: varying mean surface pressure, $p_s$, and mean surface longwave optical depth, $\tau$, between 0.01 bar and 5 bar\footnote{MarsWRF requires $p_s$ to be multiples of modern Mars pressure (610~Pa). Here we choose the multiplier to be 2, 17, 167, 833, which correspond to $p_s$ being 0.012, 0.103, 1.02, 5.08~bar.} and between 0.003 and 5, respectively. The range of values is chosen so as to compare with earlier work \cite{forget2013,kamada2021}. We note that higher values of $p_s$ or $\tau$ lead to an energy flux imbalance at the top of the atmosphere ($>5\%$ imbalance compared to the net shortwave flux) in MarsWRF. Our simulations are performed with an idealized topography, which is a Gaussian-shaped mountain placed at the equator (blue dashed contours in Fig.~1a $\&$ Fig.~1b):

\begin{equation}
    Z_s = 6000 \times e^{-\frac{1}{2}\frac{X^2}{9^2}}\times e^{-\frac{1}{2}\frac{Y^2}{7^2}}
\end{equation}
where $Z_s$ is surface elevation (in meters), and $X$ and $Y$ are longitude and latitude grid points ($-35.5\leq X \leq 35.5$, $-17.5 \leq Y \leq 17.5$), respectively.

\justifying
The surface sensible heat flux is given by:

\begin{equation}
    SH = \rho c_p C_h U^* (\theta_a - \theta_s)
\end{equation}
where $\rho$ is near-surface air density, $c_p$ is the specific heat capacity of air, $C_h$ is a heat exchange coefficient, $U^*$ is friction velocity, $\theta_a$ is near-surface potential temperature, and $\theta_s$ is surface potential temperature, respectively. Both $C_h$ and $U^*$ are calculated inside MarsWRF's surface layer scheme \cite{zhang1982}, which uses Monin-Obukhov similarity and accounts for four stability categories: stable, mechanically induced turbulence, unstable forced convection, and unstable free convection.

\justifying
To connect our idealized simulations to more Mars-relevant scenarios, we perform the following sensitivity tests: (1) obliquity set to $20^\circ$; (2) obliquity set to $20^\circ$ and with condensation and sublimation of ice caps \cite{chow2019flow} (no atmospheric collapse is found in this case); and (3) modern Mars topography. For each set of sensitivity tests, we vary $\tau$ while fixing $p_s$ to 1~bar, and we vary $p_s$ and while fixing $\tau$ to 0.1. 

\justifying
We also perform two sets of mechanism-denial experiments to verify the role of sensible heat flux, $SH$, as well as the role of atmospheric mass in modulating $SH$. In the first set of mechanism-denial experiments, $SH$ is forced to be 0. In the second set of mechanism-denial experiments, the value of $\rho$ in Eq.~(3) is held fixed at the reference value for a 1-bar atmosphere, thereby eliminating the direct effect of surface pressure on the surface turbulent heat flux. The effect of varying atmospheric mass is still considered in all other components of the model, and $SH$ is allowed to change as a result of indirect effects.

\subsection{Definition of the orographic temperature control: relative surface lapse rate, $\gamma$}
\justifying
The relationship between surface temperature, $T_s$, and elevation, $Z_s$, is quantified via the surface lapse rate, $\Gamma_s$:

\begin{equation}
    \Gamma_s = -\frac{dT_s}{dZ_s}
\end{equation}
where $dT_s/dZ_s$ is quantified by calculating a linear regression of the time-mean model output in the tropical belt $20^\circ$ N - $20^\circ$ S (see the red dashed lines in Fig.~1a and white dashed lines in Fig.~1b). We also analyzed the effect of topography in the mid-latitudes (Supplementary Information~B) to test the sensitivity to the choice of latitude.

\justifying
Furthermore, we define the relative surface lapse rate, $\gamma$, as the surface lapse rate scaled by the atmospheric dry adiabat:

\begin{equation}
    \gamma = \frac{\Gamma_s}{\Gamma_a} \times 100 \%
\end{equation}
Thus we expect, for the Earth-like regime (``mountaintops are cold"), $\gamma$ to be close to $100 \%$.

\section{Results}
\subsection{The transition of surface lapse rate}

\begin{figure}[h]
    \flushleft
    \includegraphics[width=150mm]{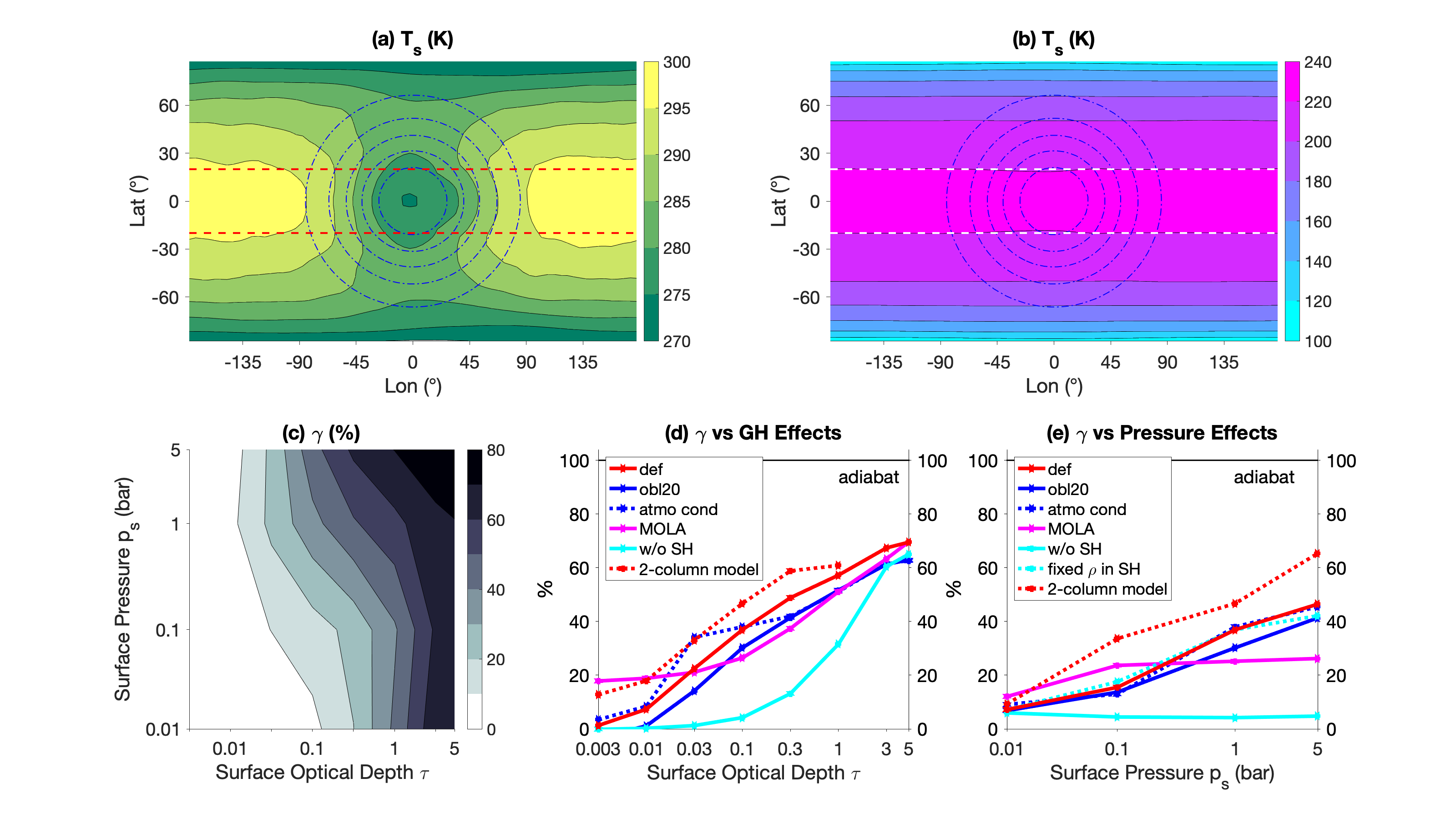}
    \caption{(a) Surface temperature $T_s$ (filled contours) for surface pressure $p_s$ = 5 bar and global mean surface optical depth $\tau$ = 5. The topography is plotted in blue dashed lines with a contour interval of 1000~m from 1000~m to 5000~m. The horizontal red dashed lines indicate the zone for tropical averaging (see Section~2.2). (b) Same as (a), but for the case with $p_s$ = 0.01 bar and $\tau$ = 0.01. The horizontal white dashed lines indicate the tropical averaging zone. (c) Relative surface lapse rate, $\gamma$ (defined in Eq.~5), as a function of greenhouse effect, $\tau$, and atmospheric thickness, $p_s$. The data is sampled on a log-scale grid with $\tau = 0.003, 0.01, 0.1, 0.3, 1, 3, 5$ and $p_s = 0.01, 0.1, 1, 5$ bar. (d) The dependence of $\gamma$ on $\tau$ when $p_s = 1$~bar. Red solid (def): default simulation, obliquity equals zero, no atmospheric condensation, idealized topography, sensible heat flux enabled. Blue solid (obl20): as def, but with obliquity set to $20^\circ$. Blue dotted (atmo cond): as def, but with obliquity set to $20^\circ$, and a $\mathrm{CO_2}$-like atmospheric condensation is enabled. Magenta (MOLA): as def, but with Mars Orbiter Laser Altimeter topography. Cyan solid (w/o SH): as def, but with sensible heat flux disabled. Red dotted (2-column model): calculations from the simple two-column model (see Section~3.4). (e) As (d), but with varying $p_s$ and fixed $\tau = 0.1$. Cyan dotted (fixed $\rho$ in SH): as def, but the value of air density, $\rho$, is held fixed at the reference value for a 1-bar atmosphere in Eq.~(3).}
\end{figure}

\justifying
We first examine the horizontal distribution of temperature in our GCM simulations. Fig.~1a$\&$1b show the typical annual mean surface temperature, $T_s$, in different climates. In all simulations, $T_s$ decreases with increasing latitude. This is consistent with our default setting of obliquity to $0^\circ$ (polar cold traps are created by radiation deficits). The pattern of near-surface atmospheric temperature, $T_a$, follows $T_s$ closely (Supplementary Information~C), with minor modulation by the winds across the elevated topography \cite{Wordsworth2015}.

\justifying
For the topographic control on surface temperature, we find two opposing limits for thick and thin atmospheres. In the thick atmosphere limit ($p_s = 5$~bar, $\tau = 5$), we find ``mountaintops are cold" (Fig.~1a): $T_s$ decreases with $Z_s$ ($\gamma \rightarrow 100 \%$). In the thin atmosphere limit ($p_s = 0.01$~bar, $\tau = 0.01$), the $T_s$ distribution becomes zonally banded (Fig.~1b), with almost no dependence on topography ($\gamma \rightarrow 0$).

\justifying
The transition of the tropical surface temperature distribution across different climates can be quantified as the change in the relative surface lapse rate, $\gamma$, with varying surface longwave optical depth, $\tau$, and surface pressure, $p_s$ (Fig.~1c). We find $\gamma$ increases with $\tau$ and $p_s$. However, the role of the greenhouse effect (i.e., variations in $\tau$) and the pressure effect (variations in $p_s$) are not symmetric. Within our parameter space, we find $\gamma$ always increases significantly with $\tau$ for any given $p_s$, but $\gamma$ increases significantly with $p_s$ only for intermediate values of $\tau$ ($\tau\sim 0.1$). When $\tau \leq 0.01$, we find $\gamma \approx 0$ for all values of $p_s$. When $\tau > 1$, $\gamma$ is close to saturation and increases only slowly with $p_s$. Sensitivity of $\gamma$ on $p_s$ is smaller than that on $\tau$ even at intermediate $\tau$. For example, starting from $\tau = 0.1$, $p_s = 1$~bar (i.e., a cold early Mars), decreasing $\tau$ by one order of magnitude leads to $\gamma$ decreasing from 37$\%$ to 7$\%$ (red solid line in Fig.~1d), while decreasing $p_s$ by one order of magnitude leads to $\gamma$ decreasing from 37$\%$ to 15$\%$ (red solid line in Fig.~1e).

\justifying
We also explore the change of $\gamma$ with $\tau$ and $p_s$ in sensitivity tests. Here we focus on Mars-relevant scenarios when the obliquity is non-zero, atmospheric condensation occurs, or the topography is different. We find $\gamma$ is slightly smaller, but still increases with $\tau$ and $p_s$, when the obliquity is non-zero (compare the blue solid line to the red solid line in Fig.~1d$\&$1e). The relationship between $\gamma$ and $\tau$ and $p_s$ also holds when the atmosphere partially condenses and sublimates seasonally (blue dotted lines in Fig.~1d$\&$1e). Changing to Mars topography decreases the sensitivity of $\gamma$ on $\tau$ and, in particular, $p_s$, although the qualitative results remain robust (magenta lines in Fig.~1d$\&$1e, also see Supplementary Information~A). Especially, the $p_s$ sensitivity becomes very small. The surface lapse rate in the mid-latitudes (discussed in Supplementary Information~B) ) also shows similar sensitivity to $\tau$, but is virtually insensitive to $p_s$. In conclusion, we confirm that both the greenhouse effect and the pressure effect can contribute to the lapse rate transition, as suggested by earlier studies \cite{forget2013,Wordsworth2016}, but the greenhouse effect dominates and is more robust.

\subsection{Surface energy budgets}
\justifying
How do the longwave optical depth and surface pressure control the surface lapse rate in different climates? \citeA{Wordsworth2016} proposed using the surface energy budget to understand the mechanisms controlling the surface lapse rate. The surface energy budget is:

\begin{equation}
    SW + LW_a = LW_s + SH
\end{equation}
where $SW$ is the net shortwave heating from the star, $LW_a$ is the longwave heating from the atmosphere (greenhouse effect), and $LW_s$ is the longwave cooling by surface emission, which is directly related to $T_s$:

\begin{equation}
    LW_s = \sigma T_s^4
\end{equation}
where $\sigma = 5.67\times 10^{-8}\; \mathrm{W/m^2/K^4}$ is the Stefan-Boltzmann constant. $SH$ is the sensible heat flux. There is no latent heat term in the surface energy budget because water vapor and $\mathrm{CO_2}$ condensation are disabled in our default simulations.

\justifying
To better understand the role of atmospheric optical thickness and pressure on the surface temperature structure, we analyze the surface energy budget in our simulations with varying $\tau$ and $p_s$. To visualize the model output, we apply a tropical meridional average ($20^\circ$ N - $20^\circ$ S) and time-average to the model output. With this approach, the temperature gradient due to solar insolation is minimized. The annual mean longitudinal variation in $T_s$ corresponds to the $LW_s$ term (see Eq.~7). For example, the red line in Fig.~2c indicates a temperature minimum at longitude = $0^\circ$ in the tropics, which corresponds to the highland in our idealized topography (Fig.~1a).

\begin{figure}[h]
\includegraphics[width=150mm]{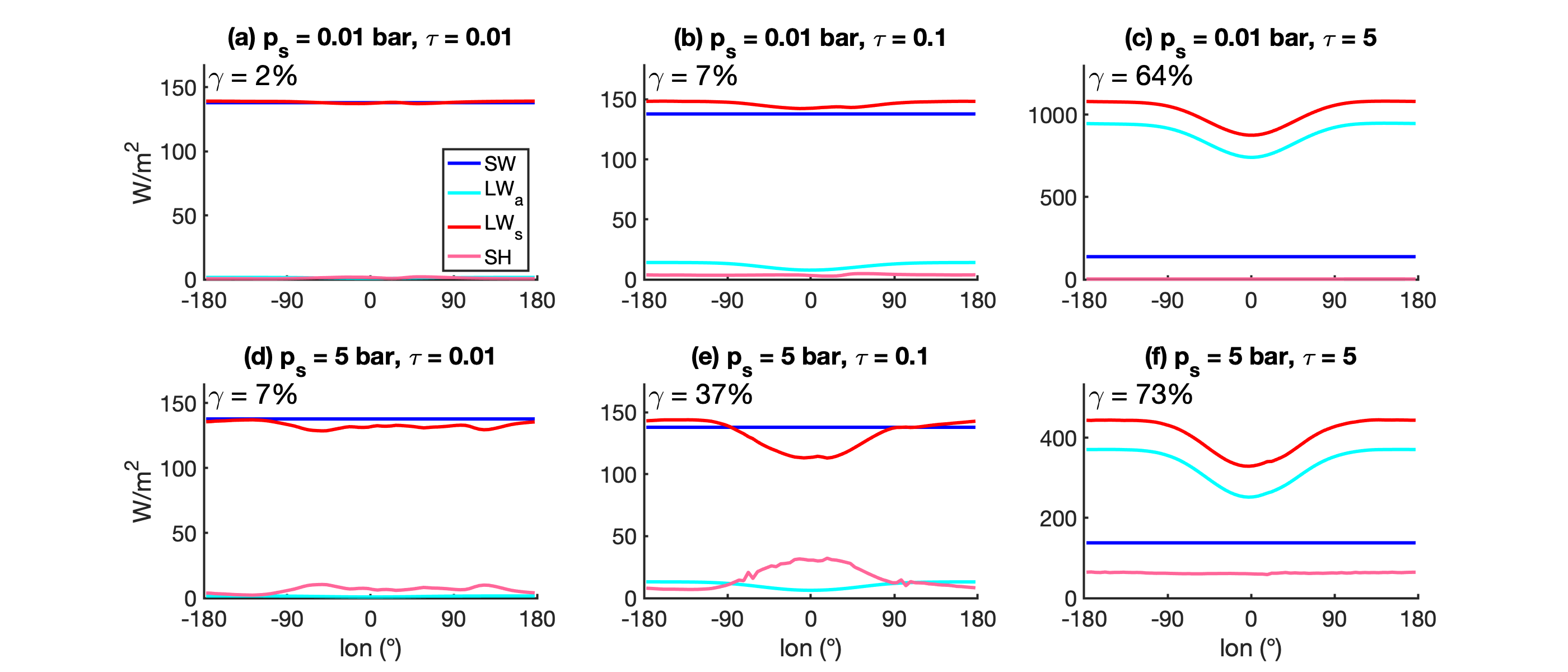}
\caption{Time-averaged surface energy budgets for typical scenarios: (a) $p_s$ = 0.01 bar, $\tau$ = 0.01, (b) $p_s$ = 0.01 bar, $\tau$ = 0.1, (c) $p_s$ = 0.01 bar, $\tau$ = 5, (d) $p_s$ = 5 bar, $\tau$ = 0.01, (e) $p_s$ = 5 bar, $\tau$ = 0.1, (f) $p_s$ = 5 bar, $\tau$ = 5. Relative surface lapse rate (surface lapse rate scaled by adiabat), $\gamma$, for each case is indicated in the upper-left corner. $SW$ is the net shortwave heating from the star, $LW_a$ is the longwave heating from the greenhouse effect, $LW_s$ is the longwave cooling by surface emission, and $SH$ is the cooling by sensible heat flux, respectively. Each term is meridionally averaged within the tropics ($20^\circ$N - $20^\circ$S). A dip in the red curve indicates a correlation between $T_s$ and topography (lower $T_s$, thus lower emission over the highlands - see Eq.~7).}
\end{figure}

\justifying
We find three typical scenarios for the zonal structure of the tropical surface energy budget (Fig.~2, see Supplementary Information~D for all cases). For an atmosphere that is optically transparent ($\tau < 0.1$, Fig.~2a$\&$2d), or thin and optically intermediate ($p_s\leq 0.1$~bar, $\tau \sim 0.1$, Fig.~2b), the major balance is between surface emission ($LW_s$, red lines) and shortwave absorption ($SW$, blue lines). The other terms are small. Since $SW$ does not vary with topographic elevation in our model setup, $LW_s$ (and thus $T_s$) can't vary much either. Hence, $\gamma$ is close to zero under this scenario. For optically thick atmospheres ($\tau \geq 1$, Fig.~2c$\&$2f), the dominant balance is between surface emission, $LW_s$, and longwave heating ($LW_a$, cyan lines). The longwave heating is weaker over the highlands compared to the lowlands, thus the highland surface is colder. Although the magnitude of $SH$ is non-negligible for massive atmospheres ($p_s \leq 1$ bar), the spatial variations of $SH$ are small under this scenario. This is consistent with our earlier results that $\gamma$ is dominated by $\tau$ when $\tau \geq 1$ (Fig.~1c). For massive, moderately opaque atmospheres ($p_s \geq 1$ bar, $\tau \sim 0.1$, Fig.~1e), $SH$ variations are large enough to generate a significant pattern in $LW_s$ (and thence $T_s$), while $LW_a$ is still small. Therefore, in this regime, variations in $SH$ are important for the surface lapse rate.

\begin{figure}[h]
\includegraphics[width=150mm]{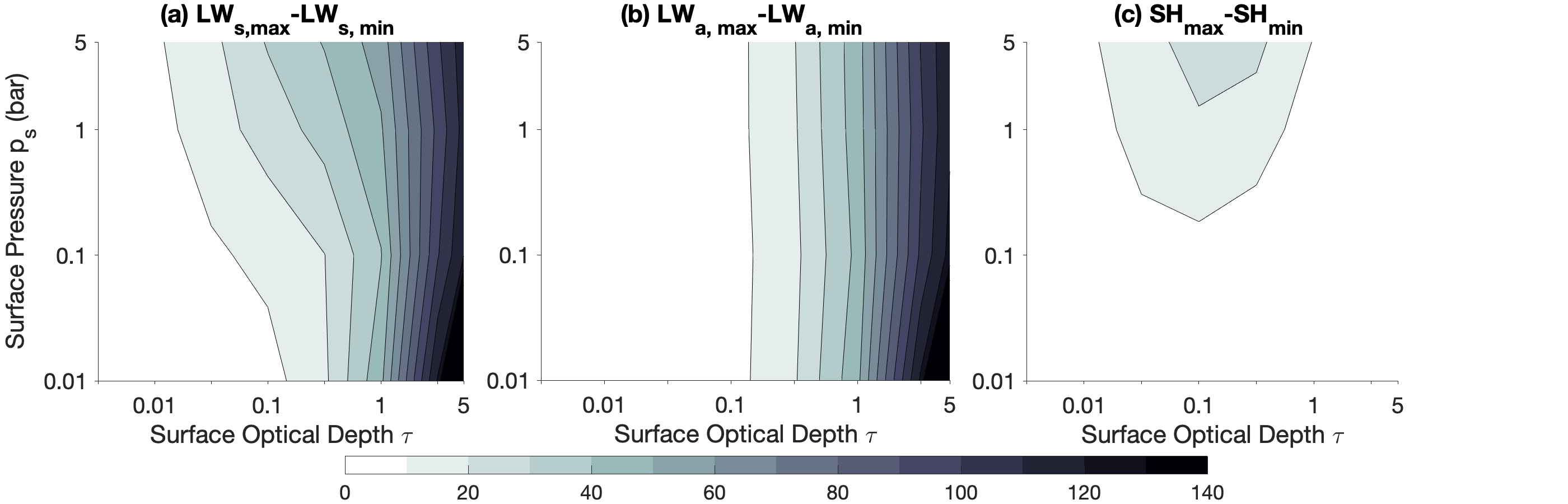}
\caption{Time-averaged surface energy budgets (unit in $\mathrm{W/m^2}$) for all cases in Fig.~1c. Same as Fig.~2, each variable is meridionally averaged within the tropics ($20^\circ$N - $20^\circ$S). (a) Contrast of surface emission, $LW_s$, (b) contrast of greenhouse effect, $LW_a$, (c) contrast of sensible heat flux, $SH$.}
\end{figure}

\justifying
To illustrate the role of spatial variations in the longwave radiation and surface heat flux across the full parameter regime, Fig.~3 shows the highland-lowland contrast as a function of $\tau$ and $p_s$ in our default simulations. Here, we focus on the difference between the maximum and minimum values (positive-definite) to visualize the highland-lowland contrast. By definition, the change in surface emission contrast (Fig.~3a) reflects the surface lapse rate transition (Fig.~1c). The surface emission contrast $LW_{s,max} - LW_{s,min}$ is balanced by contrasts in the greenhouse forcing, $LW_a$, and sensible heat flux, $SH$. The change in the $LW_a$ contrast follows $\tau$ and dominates the transition (Fig.~3b). The $SH$ contrast additionally contributes to the surface energy budget pattern when the atmosphere is moderately opaque (Fig.~3c). 

\justifying
\citeA{Wordsworth2016} proposed that $p_s$ affects $\gamma$ by modulating $\rho$ in the equation of $SH$ (Eq.~3). To test the importance of $SH$ in general, and the role of $\rho$ in the surface heat exchange equation in particular, we perform two sets of mechanism-denial experiments with modified $SH$. In the first set of simulations, we manually disable $SH$, which significantly reduces $\gamma$ when the atmosphere is intermediately opaque ($\tau \sim 0.1$), such that $\gamma$ remains small at all surface pressures (solid cyan lines in Fig.~1d$\&$1e). However, $\gamma$ still becomes large for very large $\tau$. In the second set of simulations, we fix $\rho$ at the reference value for a 1-bar atmosphere in the equation for $SH$ (Eq.~3), which does not significantly change the sensitivity of $\gamma$ to $p_s$ (dotted cyan line in Fig.~1e). Thus, $SH$ is necessary for the increase of $\gamma$ with $p_s$ at intermediate $\tau$, but the increase of 
$\rho$ in Eq.~(3) is not the primary mechanism.

\subsection{Surface lapse rate transition in a two-column model}
\justifying
To improve our understanding of the mechanisms that cause the surface lapse-rate transition, we construct a simpler two-column model, building on previous work by \citeA{yang2014}, \citeA{wordsworth2015atmospheric}, and \citeA{koll2016}. Similar to \citeA{yang2014}, our model is formulated by requiring energy balance in the free troposphere and at the surface (Eq.~8 - Eq.~11), making the Weak Temperature Gradient (WTG) approximation (Eq.~12), and enforcing convective stability or neutrality (Eq.~13 - Eq.~14). The model is therefore built on the hypothesis that the key ingredients to explain the surface lapse rate include 1) a WTG in the free troposphere, 2) convection, which maintains an adiabatic lapse-rate if and only if the radiative-advective equilibrium solution is unstable, and 3) radiative-advective equilibrium (i.e. negligible turbulent heat flux) when the solution is stable. The model equations are:

\begin{equation}
    SW - F_{c,HL} + \epsilon_{HL} \sigma T_{a,HL}^4 - \sigma T_{s,HL}^4 = 0   
\end{equation}
\begin{equation}
    F_{c,HL} - F_a  - F_H + \epsilon_{HL} \sigma T_{s,HL}^4  - 2 \epsilon_{HL} \sigma T_{a,HL}^4 = 0    
\end{equation}
\begin{equation}
    F_{c,LL} + \frac{\alpha}{1-\alpha} F_a  - F_H + \epsilon_{LL} \sigma T_{s,LL}^4  - 2 \epsilon_{LL} \sigma T_{a,LL}^4 = 0    
\end{equation}
\begin{equation}
    SW - F_{c,LL} + \epsilon_{LL} \sigma T_{a,LL}^4 - \sigma T_{s,LL}^4 = 0   
\end{equation}
\begin{equation}
    T_{a,HL} - T_{a,LL} = 0  
\end{equation}
\begin{equation}
DSE_{s,HL} \begin{cases}
= DSE_{a,HL} &\text{convective highland, solve for $F_{c,HL}$}\\
< DSE_{a,HL} &\text{stratified highland, $F_{c, HL} = 0$}
\end{cases}
\end{equation}
\begin{equation}
DSE_{s,LL} \begin{cases}
= DSE_{a,LL} &\text{convective lowland, solve for $F_{c,LL}$}\\
< DSE_{a,LL} &\text{stratified lowland, $F_{c,LL} = 0$}
\end{cases}
\end{equation}
Here, $T_{s,HL}$ and $T_{s,LL}$ are the surface temperatures of the highland and lowland, respectively; $T_{a,HL}$ and $T_{a,LL}$ are the free-tropospheric temperature of the highland and lowland, respectively; $F_{c,HL}$ and $F_{c,LL}$ are the convective heat flux from the surface to the free troposphere in the highland and lowland column, respectively; $F_a$ represents atmospheric heat transport between the columns; $F_H$ represents the atmospheric heat outflow from the tropical band caused by the Hadley circulation; $\alpha$ is the surface area fraction of the highland within the latitudinal belt; $\epsilon_{HL}$ and $\epsilon_{LL}$ are the atmospheric emissivities of the two columns; and $DSE$ is the dry static energy at the respective location. The values of $\epsilon$ and $DSE$ are calculated as:

\begin{equation}
    \epsilon_{HL} = \frac{p_{s,HL}}{p_s} \tau
\end{equation}
\begin{equation}
    \epsilon_{LL} =  \frac{p_{s,LL}}{p_s} \tau
\end{equation}
\begin{equation}
    DSE_{i} = c_p T_i + g Z_i 
\end{equation}

\justifying
The choices of parameters are explained in Supplementary Information~E. Using the seven equations (8)-(14), we numerically determine solutions for the seven dependent variables of the model: surface temperature ($T_{s,HL}$ and $T_{s,LL}$), atmospheric temperature ($T_{a,HL}$ and $T_{a,LL}$), atmospheric heat transport between the columns $F_a$, and convective heat flux ($F_{c,HL}$ and $F_{c,LL}$). We note that the model can be further simplified by combining Eq.~(9),~(10) and~(12) into a single equation to eliminate $F_a$ and merge $T_{a,HL}$ and $T_{a,LL}$ into a single unknown ($T_{a,HL}=T_{a,LL} \equiv T_a$). Moreover, since convection is never active in the lowland for any of the presented solutions, we can obtain the same results by setting $F_{c,LL}=0$ and eliminating Eq.~(14). However, we found that these simplifications provide no additional insight, and numerical solution is trivial with either formulation. Similar to \citeA{wordsworth2015atmospheric}, our model is constructed for cases when $\tau \ll 1$. For optically thick atmospheres ($\tau > 1$), the surface is no longer radiatively heated by the same atmospheric layer that emits to space, thus the single-layer-atmosphere approximation breaks down. 

\begin{figure}[h]
\includegraphics[width=140mm]{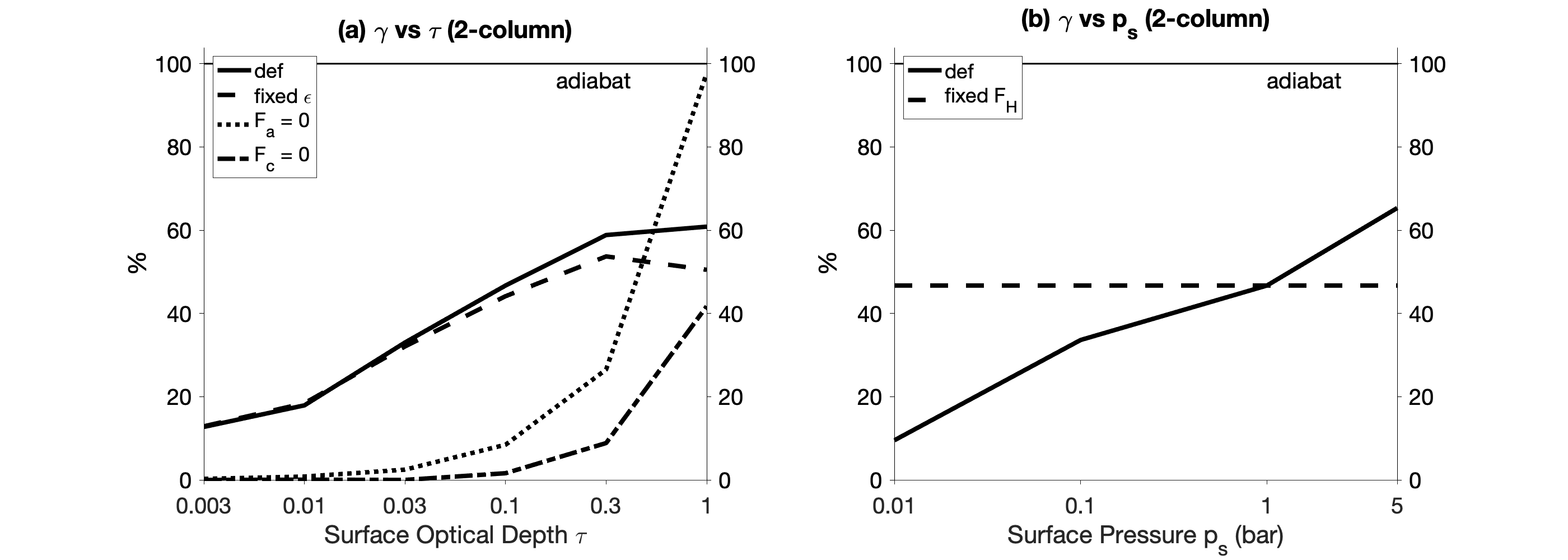}
\caption{The transition of relative lapse rate, $\gamma$, in the two-column model. (a) The dependence of $\gamma$ on the greenhouse effect, $\tau$. Solid line (def): default case - same as the red dotted line in Fig.~1d. Dashed line (fixed $\epsilon$): the case with uniform greenhouse forcing above highlands and lowlands. Dotted line ($F_a = 0$): the case with no heat advection between the highland and lowland atmosphere. Dash-dotted line ($F_c = 0$): the case with no convection between the atmosphere and surface. (b) The dependence of $\gamma$ on the pressure effect. Solid line (def): default case - same as the red dotted line in Fig.~1e. Dashed line (fixed $F_H$): the case with meridional heat advection fixed to the 1~bar value ($F_H = 6.5528$ W/m$^2$).}
\end{figure}

\justifying
The two-column model is capable of reproducing the increase of $\gamma$ with $\tau$ and $p_s$ (red dotted lines, Fig.~1d$\&$1e). From the surface energy budget perspective, our two-column model is qualitatively consistent with the transitions in the highland-lowland contrast from the GCM (Supplementary Information~E).

\justifying
In addition to confirming that the assumptions entering the two-column model formulation appear to be sufficient to understand the lapse rate transition, the model provides some insight into the specific mechanisms. The surface energy budget analysis discussed above indicated that differential longwave radiation between the lowland and highland is important to understand the surface lapse rate at high optical thickness. A naive interpretation is that $LW_a$ is smaller over the highland simply because the overlying atmosphere is less massive and hence has a weaker greenhouse effect. We can test this hypothesis in the two column model by eliminating the difference in $\epsilon$ between the two columns (setting $\frac{p_{s,HL}}{p_s} = \frac{p_{s,LL}}{p_s} = 1$ in Eq.~15$\&$16). We find the change of $\gamma$ with $\tau$ persists in this sensitivity experiment, and indeed is only weakly affected (compare solid and dashed lines in Fig.~4a).

\justifying
So what instead explains the transition with $\tau$? In the two-column model, the increase of the surface lapse rate, $\gamma$, is directly related to the lapse rate in the lowland column, due to the assumption of WTG in the atmosphere and convective adjustment over the highlands. The importance of these two assumptions can be illustrated by setting either $F_a$ or $F_c$ to zero (by modifying Eq.~12 or Eq.~13), which changes the relationship between $\gamma$ and $\tau$ significantly (dotted line and dash-dotted line in Fig.~4a). Notably, without convective heat transport between the surface and atmosphere, the sensitivity of $\gamma$ effectively reproduces the GCM simulations without $SH$ (solid cyan line in Fig.~1d). WTG and convective adjustment link the surface lapse rate to the lowland atmospheric lapse rate, which, in turn, is governed by radiative-advective equilibrium. As discussed in \citeA{payne2015} and \citeA{cronin2016}, the lapse rate of a column in radiative-advective equilibrium increases under increasing greenhouse forcing, as the increased radiative flux cools the atmosphere and heats the surface.  

In the next step we examine the sensitivity to the global mean surface pressure, $p_s$. In the two-column model, $p_s$, by construction, affects the solution only via its indirect effect on the meridional heat transport $F_H$, which is here diagnosed from the GCM simulations. The sensible heat flux, $SH$, which in the real world is associated with complex boundary layer physics \cite{joshi2020}, is implied by Eq.~13$\&$14. Consistent with \citeA{kaspi2015}, we find that higher $p_s$ (and thus greater atmospheric mass) drives larger $F_H$. Following our argument above, atmospheric heat flux divergence from the tropics, $F_H$, leads to a reduction of the net atmospheric heat flux convergence over the lowlands, which leads to a reduction in the atmospheric lapse rate (and, hence, a reduction in the surface lapse rate, $\gamma$). The mechanism can be illustrated by fixing $F_H$. As expected, we find that when $F_H$ is fixed, the sensitivity of $\gamma$ to $p_s$ disappears (Fig.~4b).

Taken together with the GCM simulations, the results suggest that both a weak temperature gradient and highland convection are important for explaining the sensitivity of $\gamma$ to the greenhouse effect and pressure effect. Meanwhile, the spatial variations in the column-integrated greenhouse gas and the near-surface air density do not play a major role in the sensitivity of the surface lapse rate.

\section{Discussion and Summary}

\justifying
``Radiation deficits are cold" and ``mountaintops are cold" comprise the usual expectation for the distribution of surface temperature, $T_s$. Here, using a GCM simulating a fast-rotating, dry planet, we argue that whether ``mountaintops are cold" depends on both the greenhouse effect (longwave optical depth, $\tau$) and surface pressure, $p_s$. Specifically, the dependence of $T_s$ on surface elevation is quantified as the tropical surface lapse rate relative to the adiabat, $\gamma$. We find that $\gamma$ increases with $\tau$ and $p_s$. However, the roles of the greenhouse effect and pressure effect are not symmetric. At all surface pressures, we find that $\gamma$ is close to zero for very small $\tau$, and approaches 100$\%$ as $\tau \gg 1$. Surface pressure plays a significant role at intermediate $\tau$, where more massive atmospheres tend to have larger $\gamma$ in the tropics, but the effect of $p_s$ is less robust. From a surface energy budget perspective, spatial variations in the downwelling atmospheric longwave radiation $LW_a$ balance the variations in the upward longwave radiation associated with the topographic surface temperature variations for optically thick atmospheres ($\tau > 1$). For optically moderate atmospheres ($\tau \sim 0.1$) the variations in surface longwave emission $LW_s$, associated with the topographic temperature variations, are maintained by variations in the sensible heat flux, $SH$. Large $\gamma$ requires a weak temperature gradient in the atmosphere and effective coupling between the surface and the atmospheric temperature, where the coupling can occur either radiatively or via $SH$. The surface lapse rate transition can be reproduced in a two-column, two-layer model, consisting of a convective highland column together with a stable lowland column, coupled via the weak temperature gradient assumption in the atmosphere. The two-column model suggests that weak temperature gradient and highland convection are important to explain the lapse rate transition. Increases in optical thickness or surface pressure then affect the tropical surface lapse rate by destabilizing the atmospheric lapse rate over the lowlands. 

\justifying
This paper focuses on the surface temperature distribution on fast-rotating, dry planets. We can speculate how the conclusions might differ on other planets, although future work should use GCMs to verify these predictions. (1) For tidally locked planets, solar insolation never reaches the permanent night hemisphere, such that the nightside, heated by advection, is stably stratified \cite{joshi2020,Ding2021}. Thus, the mountaintops in the night hemisphere might be warm depending on the strength of the thermal inversion. (2) For warm, wet planets, water vapor modulates the atmospheric lapse rate from a dry adiabat towards the moist adiabat, which is likely to similarly affect the surface lapse rate for thick atmospheres. (3) For cold, wet planets, the ice-albedo feedback is important. The existence of ice would decrease absorbed solar insolation. In the optically thin limit, one would then expect to find a surface temperature discontinuity near the snowline. In the optically thick limit, our mechanism suggests that the surface lapse rate still approaches the adiabat, independent of the presence of snow or ice (which is broadly consistent with present-day Earth). (4) The influence of spectral properties of real gas species are ignored in this work. For real gas species, surface pressure and greenhouse effect are coupled due to the effect of pressure broadening and collision-induced absorption (CIA) \cite{ding2019new}. Meanwhile, for optically thick atmospheres with typical greenhouse species (e.g., $\mathrm{CO_2}$, $\mathrm{H_2O}$), the surface can still emit to space through the spectral windows. Different greenhouse gas emission spectra are therefore likely to affect the quantitative results, although we have found that changing optical thickness in our gray atmosphere qualitatively reproduces the effect of increasing $\mathrm{CO_2}$. (5) Rayleigh scattering by atmospheric molecules (which influences shortwave heating, $SW$, and is related to pressure) is ignored in this paper. It was found to be unimportant within the parameter space used in our study. But the role of reflection by clouds remains unknown.

\justifying
Inspired by earlier research on Mars \cite{forget2013,wordsworth2013,Wordsworth2016,Kite2019}, our work suggests that early Martian sedimentary geology might be explained by a new end-member option for the climate: ``high $T_s$ + non-$\mathrm{CO_2}$ greenhouse gases + low $p_{\mathrm{CO_2}}$". Under this scenario, changes in fluvial patterns may arise from non-$\mathrm{CO_2}$ greenhouse forcings rather than the loss of a $\mathrm{CO_2}$-dominated atmosphere \cite{Kite2022}. The non-$\mathrm{CO_2}$ greenhouse forcing could be water ice cloud radiative forcings \cite{urata2013,kite2021}, or $\mathrm{H_2}+\mathrm{CO_2}$ CIA \cite{wordsworth2017,turbet2021}. Future work should further test the correlation between river locations and elevation with non-$\mathrm{CO_2}$ greenhouse forcings and low atmospheric pressure ($p_s < 1$ bar).

\justifying
Our work can be connected to other planets in the solar system. Based on Akatsuki infrared measurements, recent analysis show that surface temperature on Venus depends little on insolation compared to elevation \cite{singh2019venus}. However, the elevation dependence is small compared to the adiabat. It remains unclear whether the unexpectedly small dependence is a result of physical processes \cite{esposito1984sulfur} or instrument error \cite{iwagami2018initial}. Fortunately, forthcoming missions (DAVINCI, VERITAS, and EnVision) will extend the understanding with new data on atmospheric thermal structure and surface properties. Meanwhile, modeling on Titan has found robust correlation between temperature and elevation \cite{lora2019model}. The greenhouse effect on Titan is provided by methane and $\mathrm{N_2}$-$\mathrm{N_2}$ CIA. Since the lifetime of methane is short on geological timescales, it is possible that Titan has undergone a lapse rate transition after a methane outburst event \cite{tokano2021paleoclimate}.

\justifying
With respect to the habitability of exoplanets, our work also highlights the potential role of topography in creating cold traps in optically thick atmospheres, in addition to the cold traps created by radiation deficts and atmospheric circulation \cite{Ding2020}. Most exoplanet GCMs assume no topography. However, topography allows a greater chance for water to condense. Future work could focus on different potential climate regimes under the competition of these cold traps, as well as their influences on the hydrological cycle and long-term planetary evolution (e.g., the transition between snowball and habitable climates).

\acknowledgments    
This research was funded by NASA (80NSSC20K0144+80NSSC18K1476) and NSF award AGS-2033467. We thank Robin Wordsworth and an anonymous reviewer for comments that led to an improved manuscript. We also thank Tiffany Shaw for helpful discussions. Our simulations were completed with resources provided by the University of Chicago Research Computing Center (RCC). A portion of this work was carried out at the Jet Propulsion Laboratory, California Institute of Technology, under a contract with the National Aeronautics and Space Administration (80NM0018D0004). 

\section*{Open Research}
Data necessary to reproduce the figures in this paper is publicly available on Zenodo \cite{bowen_fan_2023_8404011}.

\bibliography{agusample.bib}
\end{document}


%
%


\title{Supporting Information for \textit{Why are Mountaintops Cold? The Transition of Surface Lapse Rate on Dry Planets}}
%
%



\authors{Bowen Fan\affil{1}, Malte F. Jansen\affil{1}, Michael A. Mischna\affil{2}, Edwin S. Kite\affil{1}}

\affiliation{1}{Department of the Geophysical Sciences, University of Chicago, Chicago, IL, 60637, USA}
\affiliation{2}{Jet Propulsion Laboratory, California Institute of Technology, Pasadena, CA, 91109, USA}








\appendix
\section{Simulations with a pure $\mathrm{CO_2}$ atmosphere}

\begin{figure}[b]
\includegraphics[width=180mm]{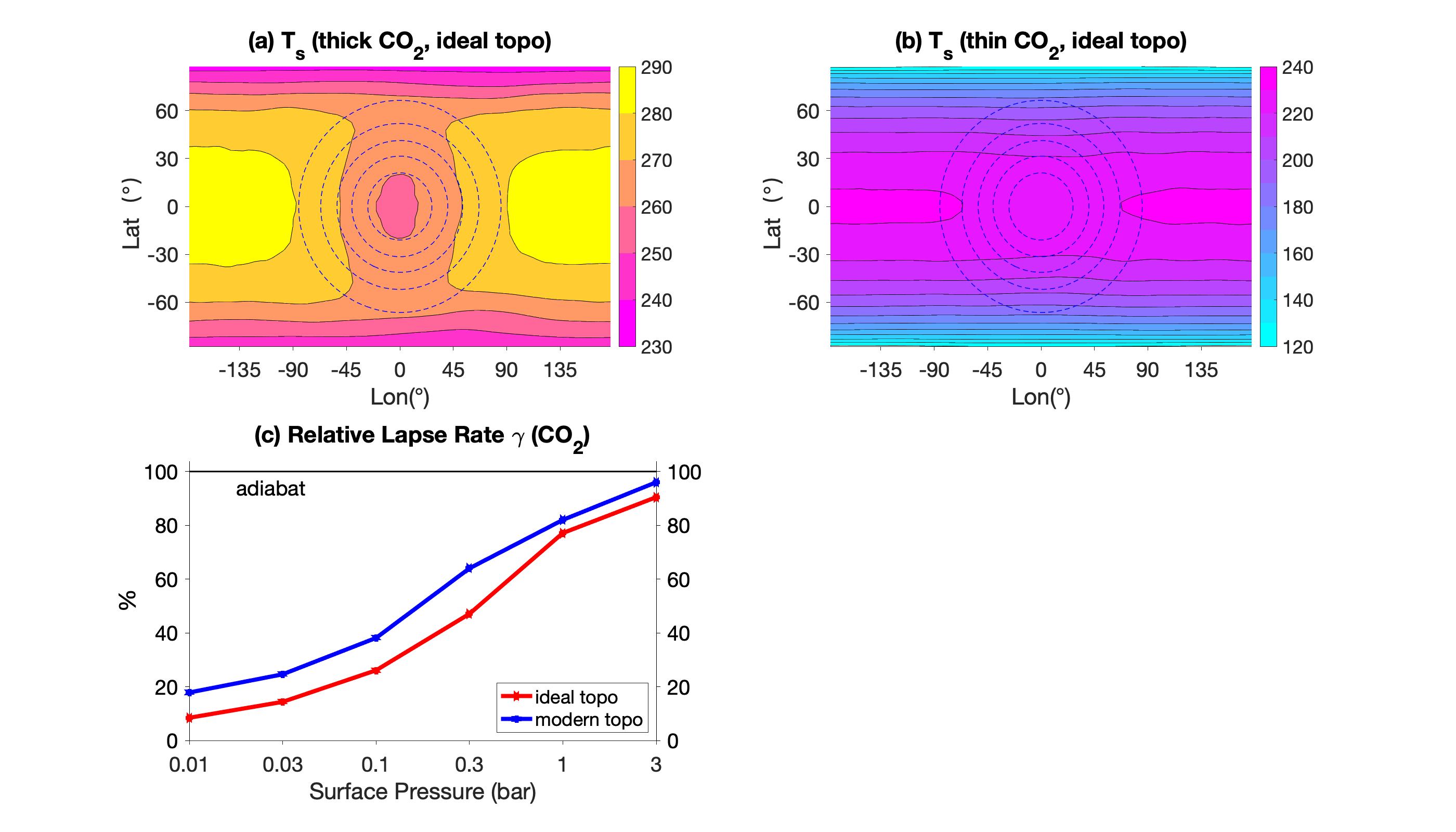}
\caption{Example of annual mean surface temperature, $T_s$, patterns and relative surface lapse rates ($\gamma$) under $\mathrm{CO_2}$ atmospheres. (a) Surface temperature (filled contours) for surface pressure $p_s$ = 3 bar. The ideal topography is plotted in blue dashed lines with a contour interval of 1000~m from 1000~m to 5000~m. (b) Same as (a), but with $p_s$ = 0.01 bar. (c) The transition of $\gamma$ with $p_s$. The data is sampled on a logarithmic grid with $p_s = 0.01, 0.03, 0.1, 0.3, 1, 3$ bar. Red solid (ideal topo): obliquity equals zero, no atmospheric condensation, default idealized topography, sensible heat flux enabled. Blue solid (modern topo): same as ideal topo, but with modern Mars topography.}
\end{figure}

\justifying
To validate our GCM results against earlier work on early Mars \cite{forget2013,Wordsworth2016}, we carried out simulations with a pure $\mathrm{CO_2}$ atmosphere using a correlated-k scheme \cite{Mischna2012}. The settings are the same as with the gray gas simulations (see Section~2), except for the radiation scheme. The surface pressure, $p_s$ (equivalent to $p_{CO2}$), is varied from 0.01 bar to 3 bar. We also performed a set of simulations with modern Mars topography.

\justifying
We find similar transitions in surface temperature, $T_s$, distribution (Fig.~S1a $\&$ Fig.~S1b) and surface lapse rate, $\gamma$ (Fig.~S1c), as in the gray gas runs. The value of $\gamma$ is close to 100$\%$ for a thick $\mathrm{CO_2}$ atmosphere (which has both a high surface pressure and high optical thickness), and converges to 0 for a thin $\mathrm{CO_2}$ atmosphere (which has low surface pressure and low optical thickness). Quantitatively, our simulations are consistent with Fig.~7 in \citeA{forget2013} in that $\gamma \approx 100\%$ for $p_s = 3$ bar, and $\gamma \approx 80\%$ for $p_s = 1$ bar. The slope of $\gamma$ does not depend on topography (compare the blue line and red line in Fig.~S1c). We also find that only with the correlated-k $\mathrm{CO_2}$ scheme, the upper limit of $\gamma$ exceeds $90\%$, while for the gray gas scheme, we have not obtained values of $\gamma$ above $80\%$ (Fig.~1). The cause for this difference between a gray gas and and $\mathrm{CO_2}$ atmosphere is beyond the scope of this investigation (see the discussion in Section~4).

\section{Role of off-equatorial topography and off-equatorial latitude band}

\begin{figure}[b]
\includegraphics[width=180mm]{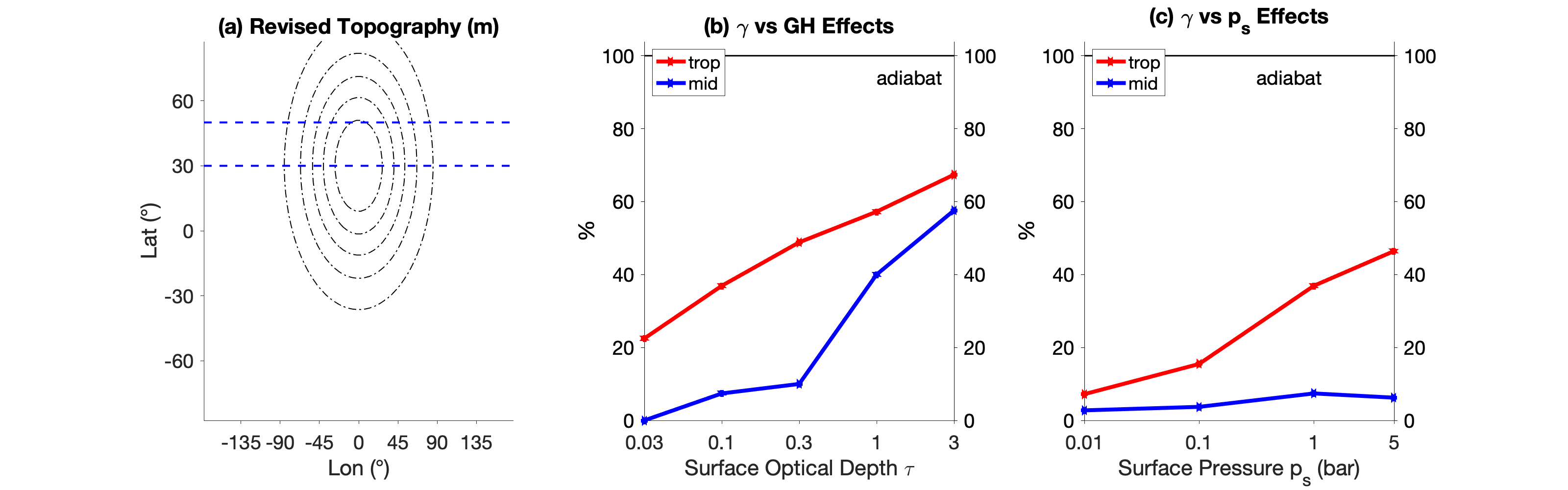}
\caption{(a) Revised Gaussian topography. The shape is the same as the original Gaussian topography, but the center of the mountain is shifted to 30$^\circ$ N. The latitudinal band for sampling data is shown in blue dashed lines. (b) The transition of surface lapse rate, $\gamma$, with $\tau$, for $p_s = 1$ bar. Red line (trop): simulations focused on the tropics, with the original Gaussian topography and data sampled from the tropical band. Blue line (mid): simulations focused on mid-latitudes, with the revised topography and mid-latitudinal band. (c) Same as (b), but for $\tau = 0.1$ and varied $p_s$.}
\end{figure}

\justifying
To test the sensitivity to the choice of latitudinal band, we check the lapse rate transition with a shifted idealized topography and latitudinal band.

Fig.~S2 shows the data with a gray gas scheme and a revised Gaussian topography. Fig.~S2(a) shows the topography (black line) and indicates the latitudinal band used for averaging: $30^\circ$N$-50^\circ$N (blue line). Fig.~S2(b)$\&$(c) compares the transition of relative surface lapse rate, $\gamma$, between the default cases with tropical topography and tropical band (red line, also see Fig.~1) and the revised topography and mid-latitude band. We find $\gamma$ still increases with $\tau$, but $\gamma$ is much smaller when the atmosphere is optically thin ($\tau<1$ for Fig.~S2b, also in Fig.~S2c). This is because the troposphere gains heat by advection from low latitudes in those cases ($F_H < 0$, see Section~3.3), which stabilizes (i.e., reduces) the atmospheric lapse rate. Importantly, we also find the mid-latitude lapse rate to be essentially independent of $p_s$. This result is consistent with the argument from Section~3.3 that an increase in atmospheric heat flux divergence (out of the tropics) with increasing $p_s$ is the key for the tropical surface lapse rate change.

\section{The near-surface atmospheric pattern of limiting cases}
\begin{figure}[b]
    \flushleft
    \includegraphics[width=180mm]{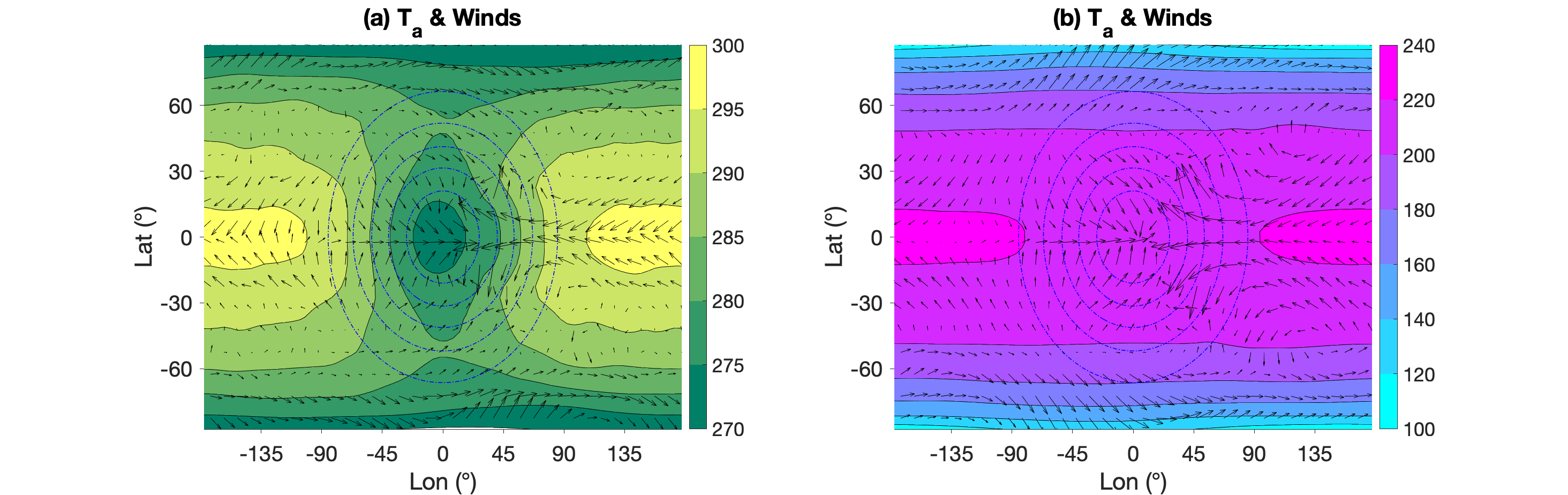}
    \caption{Example of annual mean near-surface temperature patterns under different atmospheres. (a) Near-surface atmospheric temperature $T_a$ (filled contours) and near-surface wind speeds for surface pressure $p_s$ = 5 bar and global mean surface optical depth $\tau$ = 5. The maximum zonal wind speed is 3.4 m/s and the maximum meridional wind speed is 2.5 m/s. (b) Same as (a), but for the case with $p_s$ = 0.01 bar and $\tau$ = 0.01. The maximum zonal wind speed is 10.3 m/s and the maximum meridional wind speed is 8.1 m/s.}
\end{figure}

\justifying
Fig.~S3 shows the typical patterns of near-surface atmospheric temperature, $T_a$, and horizontal winds. $T_a$ follows closely with $T_s$ (Fig.~1), with minor modulation by the winds across the elevated topography \cite{Wordsworth2015}.

\section{Full surface energy budgets for gray gas simulations}

\begin{figure}[b]
\includegraphics[width=250mm, angle=90]{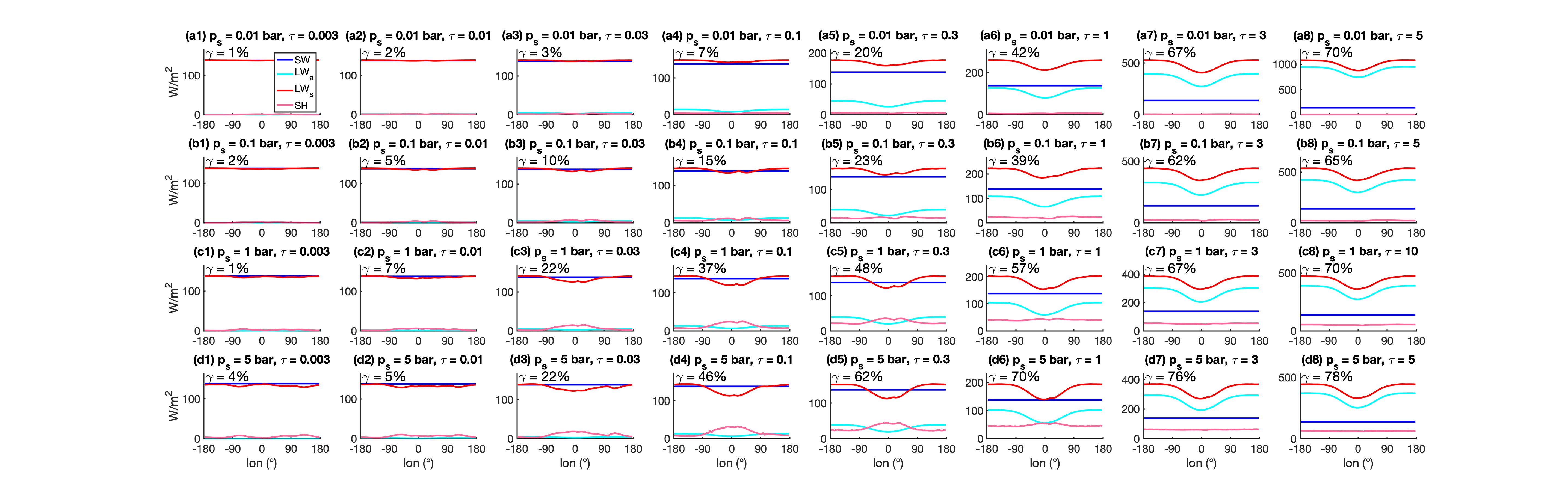}
\caption{Time-averaged surface energy budgets for all cases shown in Fig.~1c.}
\end{figure}

The energy budgets for all default gray gas simulations are shown in Fig.~S4. 

\section{Two-column model}

\justifying
This section includes the choice of parameters and the results for the two-column model. The parameters are listed in the table below. 

\justifying
The topography-related variables are calculated by spatially averaging the values within the highland box and lowland box, respectively. Here we define the highland area to include all elevations higher than the half-height of the idealized mountain ($Z_s > 3000$ m). This leads to a highland surface fraction within the tropical belt, $\alpha$, of $0.3056$. Then $Z_{s,HL}$, $Z_{s,LL}$, $p_{s,HL}$, $p_{s,LL}$ are calculated by spatial averaging over the respective regions. The calculation of the atmospheric energy budget requires an additional assumption on the atmospheric level. Here we choose the atmospheric pressure $p_c = 0.4 p_s$. At this level, the atmosphere above the highlands and lowlands has reached a weak temperature gradient in all our GCM simulations (Fig.~S5). Then the thickness of the atmospheric layer above highland $H_a$ is calculated as:

\begin{equation}
    H_a = H\frac{log \left( p_{s,HL} \right)}{log \left( p_c \right)}
\end{equation}
where $H = \frac{RT_{s,HL}}{g}$ is the scale height. The altitude of the atmosphere is calculated as $Z_a = H_a + Z_{s,HL}$.
 
\begin{figure}[b]
\includegraphics[width=230mm, angle=90]{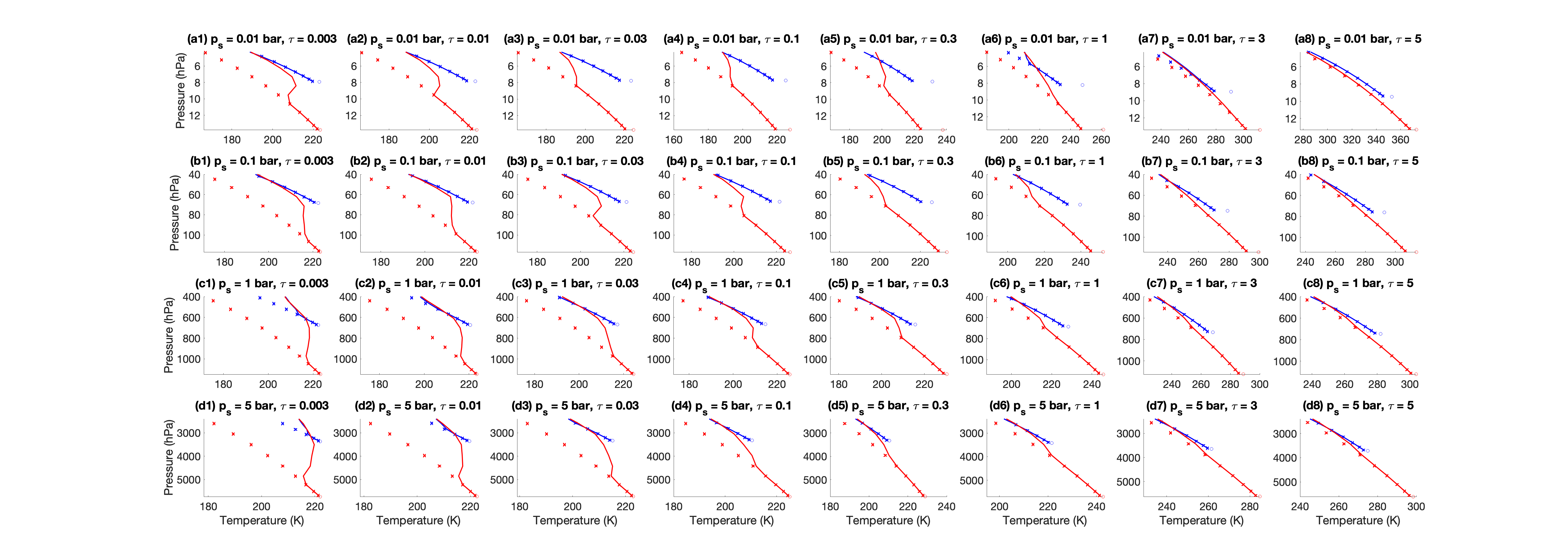}
\caption{Time-averaged vertical thermal structure above the equatorial highlands (blue, lon = $0^\circ$) and lowlands (red, lon = $180^\circ$) for all cases shown in Fig.~1c. Circles, solid lines, and crosses correspond to surface temperature, atmospheric temperature, and adiabats, respectively.}
\end{figure}

An advective cooling term, $F_H$, is included in the atmospheric equations (Eq.~10$\&$11) because the Hadley circulation transports heat from the tropics to higher latitudes. For each simulation, $F_H$ is calculated by spatially averaging the top of atmosphere (TOA) radiative imbalance within the tropical atmospheric band:

\begin{equation}
     F_H = \overline{LW_{\downarrow, TOA} + SW_{\downarrow, TOA} - LW_{\uparrow, TOA} - SW_{\uparrow, TOA}}
\end{equation}
where the overbar denotes a horizontal average, upward arrows represent fluxes that go into space, and downward arrows represent fluxes that enter the atmosphere. 

After plugging in the parameters, we numerically solve the model equations for the temperature and energy fluxes. The surface lapse rate from the two-column model is defined as:

\begin{equation}
     \gamma_{two-column\;model} = - \frac{1}{\Gamma_a}\frac{T_{s,HL} - T_{s,LL}}{Z_{s,HL} - Z_{s,LL}} \times 100\%
\end{equation}

\begin{tabular}{| c | c | c | c | c | c | c | c | c | c |}
    \hline
    insolation $SW$ ($\mathrm{W/m^2}$) & \multicolumn{9}{|c|}{137.84} \\
    \hline
    gravity $g$ ($\mathrm{m/s^2}$) & \multicolumn{9}{|c|}{3.7} \\
    \hline
    highland fraction $\alpha$ & \multicolumn{9}{|c|}{0.3056} \\
    \hline
    highland elevation $Z_{s,HL}$ (m) & \multicolumn{9}{|c|}{4873} \\
    \hline
    lowland elevation $Z_{s,LL}$ (m) & \multicolumn{9}{|c|}{754} \\
    \hline
    optical depth $\tau$ & \multicolumn{4}{|c|}{0.1} & 1 & 0.3 & 0.03 & 0.01 & 0.003 \\
    \hline
    surface pressure $p_s$ (Pa) & 5.08E5 & 1.02E5 & 1.03E4 & 1200 & \multicolumn{5}{|c|}{1.02E5} \\
    \hline
    highland pressure $p_{s,HL}$ (Pa)  & 3.79E5 & 7.60E4 & 7.68E3 & 890 & \multicolumn{5}{|c|}{7.60E4} \\
    \hline
    lowland pressure $p_{s,LL}$ (Pa) & 5.48E5 & 1.10E5 & 1.11E4 & 1290 & \multicolumn{5}{|c|}{1.10E5} \\
    \hline
    advective cooling $F_H$ ($\mathrm{W/m^2}$) & 9.6400 & 6.5528 & 4.5974 & 2.6564 & 12.0747 & 9.3281 & 4.5974 & 2.6564 & 1.8889 \\
    \hline
\end{tabular}

\begin{figure}[h]
\includegraphics[width=120mm]{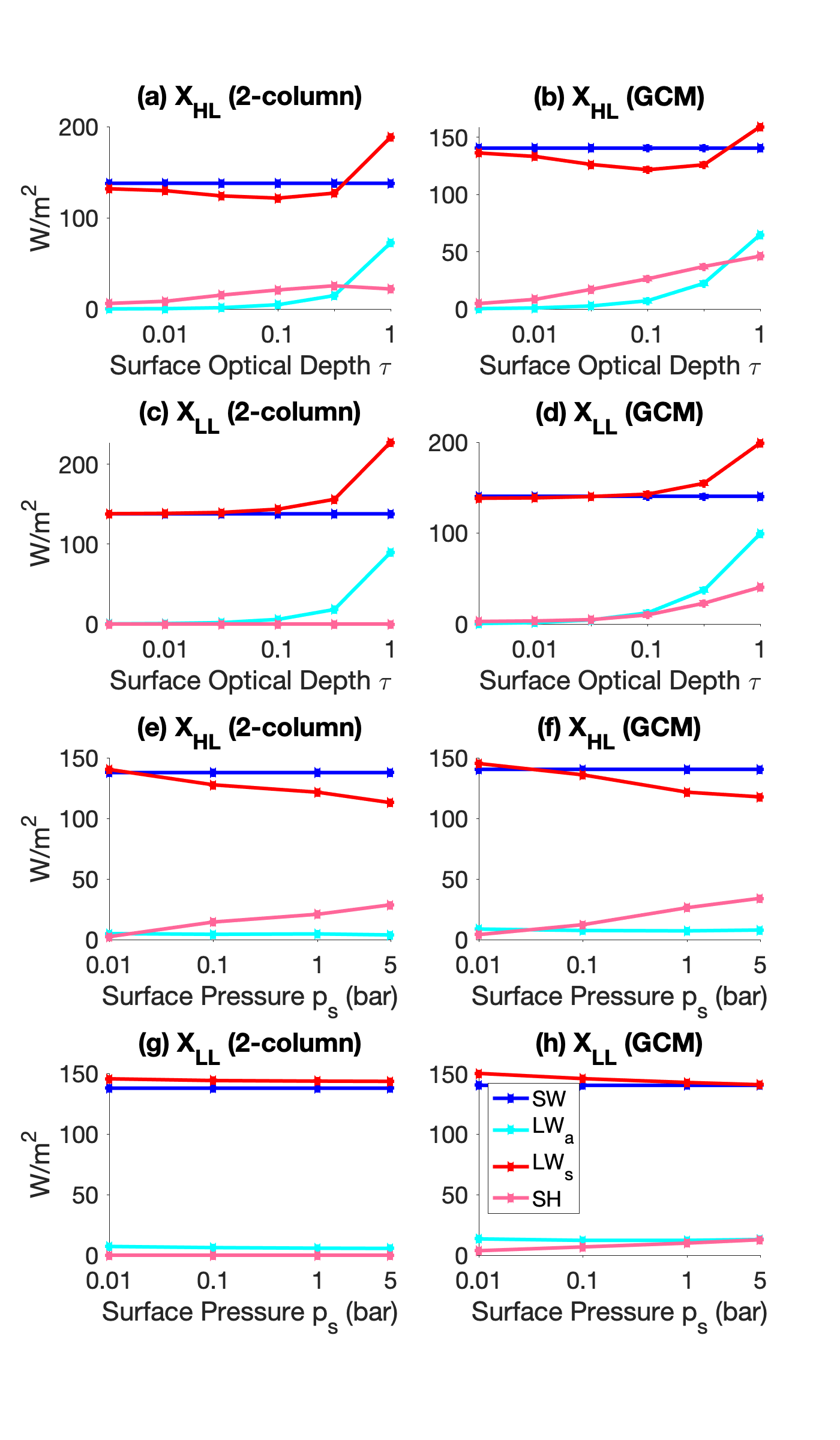} 
\caption{(a) Surface energy budget components in the highland column of the two-column model with $p_s = 1$~bar. (b) Same as (a), but for the GCM. (c) Same as (a) but for the lowland column. (d) Same as (c), but for the GCM. (e, f, g, h) Same as the top rows, but for the cases with $\tau = 0.1$ and varied $p_s$.}
\end{figure}

\bibliography{agusample.bib}